\documentstyle[prl,aps,epsfig]{revtex}

\newcommand{\beq}{\begin{equation}}
\newcommand{\eeq}{\end{equation}}
\newcommand{\beqar}{\begin{eqnarray}}
\newcommand{\eeqar}{\end{eqnarray}}
\newcommand{\bfig}{\begin{figure}}
\newcommand{\efig}{\end{figure}}

\newcommand{\bd}{\begin{itemize}} 
\newcommand{\ed}{\end{itemize}} 
\newcommand{\bc}{\begin{center}}
\newcommand{\ec}{\end{center}}
\newcommand{\be}{\begin{equation}}
\newcommand{\ee}{\end{equation}}

\newcommand{\ba}{\begin{array}}
\newcommand{\ea}{\end{array}}

\newcommand{\set}[2]{\newcommand{#1}{#2}}
\set{\pa}{\partial \over \partial\, }
\set{\leftvector}{\stackrel{\leftarrow}{\partial }}
\set{\rightvector}{\stackrel{\rightarrow}{\partial }}
\begin{document}
\twocolumn[\hsize\textwidth\columnwidth\hsize
           \csname @twocolumnfalse\endcsname

\title{Charmonium production in relativistic proton-nucleus collisions 
  : What will we learn from the negative $x_F$ region ? }
\author{
D. Koudela$^{a}$,~C. Volpe$^{a,b}$}

\address{
$^{a)}$
Institut f\"ur Theoretische Physik der Universit\"at, Philosophenweg
19, D-69120 Heidelberg, Germany \\
$^{b)}$
  Groupe de Physique Th\'{e}orique, Institut de Physique Nucl\'{e}aire,
F-91406 Orsay Cedex, France \\}

\maketitle

\begin{abstract}
We study the nuclear medium effects on the
$c\bar{c}$ time evolution and charmonium production,
in a relativistic proton-nucleus collision.
In particular, we focus on the fragmentation region of
the nucleus where the formation length of the charmonium
mesons is shorter than the size of the nucleus.  
Little is known on the nuclear effects in this region.
We use a quantum-mechanical model which includes a realistic
potential for the $c\bar{c}$ system and an imaginary potential to
describe the collisions of the $c\bar{c}$ with the nucleons. 
The imaginary potential introduces a transition amplitude among
the charmonium states and produces an interference pattern on
the charmonium survival probability, which is particulartly
important for $\psi'$.
Our results on the suppression factors are compared with data 
from the NA50 and 
E866/NuSea Collaborations. Predictions 
are given for the suppression of $J/\psi,~\psi',~\chi_c$
as a function of the nuclear mass and in the negative $x_F$ region,
where data will be available soon.
\end{abstract}

\vskip2pc
]

\noindent
A very exciting era in the study of strong
interactions has begun with the advent of high energy
heavy ion collision experiments. In particular, this has
opened the way to the exploration of new states of matter,
such as the colour glass condensate, corresponding to
the nonlinear regime of quantum chromodynamics,
or the high temperature phase
of nuclear matter, the so-called quark-gluon plasma,                        
which is also important for our understanding
of the early universe.
In relativistic heavy
ion collisions,
evidence for the formation of the
quark-gluon plasma may come from 
the combined information of different signals, such
as strangeness, dilepton, photon or charmonium production.

Matsui and Satz \cite{MSatz} first suggested that the suppression of
charmonia could  signal 
that a phase transition has occurred.
This pioneering idea 
has triggered a series of experiments
at the CERN SPS \cite{pp,fepA},
whose interpretation has been very controversial. 
It has become clear that,
to use charmonium as a signal,
all possible
mechanisms which can affect 
charmonium production need to be well understood 
\cite{GH99,Vogt}.

In this context,  proton-proton ({\it
  pp}) 
and proton-nucleus ({\it pA}) 
collisions have been intensively
studied \cite{pp,fepA,GH99,Vogt,pprecent,NA50,E866}. 
In fact, these processes are used as a reference
for the case of nucleus-nucleus ({\it AB}) collisions where 
a critical energy density  
may be eventually attained, producing 
the plasma phase. 
This should affect various observables and in particular,
it may lead to an extra (anomalous) charmonium suppression.
There have been indications recently 
that the plasma has been produced, but it
is still too early to draw definite conclusions
\cite{Blaizot}. 
These studies will be pursued
by the running experiments at RHIC, and future measurements at LHC.

It takes a certain time for a $c\bar{c}$ 
produced in a collision
to become a
colour singlet state, then it expands 
to the size of a $\psi$ meson
(the symbol $\psi$ stands from now on for 
any charmonium meson, i.e. $J/\psi,~\psi'$ and $~\chi_c$).
Different models exist at present for the production mechanisms
\cite{GH99,Kramer,Peigne}.

Sufficiently fast $c\bar{c}$ can traverse the nucleus before
a $\psi$ meson is fully formed; 
whereas slow $c\bar{c}$ can traverse the nucleus as fully
formed $\psi$ meson
\cite{KZ91,KS95}.  
Experimentally one can pinpoint to these different kinematical
regions by measuring charmonium production at positive and negative 
small Feynman $x_F$ and/or making inverse kinematics measurements 
\cite{GH99,KS95}.

One
of the aspects which need to be well understood
is the effect of the nuclear medium on the $c\bar{c}$ pair evolution. 
In particular,
very little is known on the nuclear medium effects
when the formation time of a fully developed charmonium 
is smaller than the time to traverse the nucleus \cite{KS95}.
Soon, measurements in $pA$ and $AB$ collisions will be performed by the 
HERA-B Collaboration \cite{HERAB}
at DESY and the PHENIX Collaboration at RHIC \cite{PHENIX}. Moreover
the E160 Collaboration at SLAC will perform similar experiments where
the charmonia are
photoproduced \cite{E160}.

In this letter, we focus on this region, 
i.e. the fragmentation region of the nucleus (negative $x_F$ region or slow
$c\bar{c}$) and
study
the effects of the nuclear medium 
on the $c\bar{c}$ time evolution 
 and on 
charmonium
suppression in relativistic {\it pA}  collisions \cite{DA}. 
Because of the kinematical region we are interested in,
we assume that the produced $c\bar{c}$
becomes quickly a colour singlet state (premeson) and
on its way through 
the nucleus,
this premeson expands to  
a $\psi$ meson, while experiencing  
collisions with the nucleons. To describe this,
we use here a
quantum-mechanical model 
where, contrary to previous works \cite{GH99,KZ91,KS95,Arleo}, 
 the $c\bar{c}$ pair is bound by a
realistic potential \cite{BT} and the 
premeson (and then meson) wave function is expanded on 
the basis given by the charmonium eigenstates. 
An imaginary potential depending on
the dipole charmonium-nucleon cross section \cite{KZ91,HK}
is included to describe 
the collisions with the nucleons.
This introduces a transition amplitude 
among the charmonium states.
(We assume the premeson wavefunction has a fixed
angular momentum quantum number, while it has
different radial components.)

We will show that the interaction of the premeson  and later on the meson 
with
the nuclear medium produces an interference pattern on the 
charmonium survival probability. 
We will show how this affects the charmonium suppression
both when the path in the nucleus
is varied, by changing the mass number $A$, and as a function of
$x_F$. These effects are particularly important
for $\psi'$.
We compare our results 
with experimental data
obtained at CERN SPS by the Na50 \cite{NA50} and 
at Fermilab by the
E866/NuSea \cite{E866} 
Collaborations. We present predictions for the production of 
$J/\psi,~\psi',~\chi_c$ in a range of small 
negative $x_F$ values,
for different nuclei \cite{DA}, for which experimental data will be avalaible
soon \cite{HERAB,PHENIX}. 

Our $c\bar{c}$ pair, produced in a $pA$ collision, is described
(in its center of mass frame)  by the internal Hamiltonian 
\be\label{e:1}
H_0 = {p^2 \over {m_c}} + V(r),
\ee
where $m_c$ is the charm quark mass.
The potential $V(r)$ which binds $c$ and $\bar{c}$ is a realistic
potential which 
reproduces well the properties of the charmonium family \cite{BT}.
The spin-dependent terms are neglected. 
Only transitions between charmonium states with different
$n$ quantum numbers are considered. 

By solving the static Schr\"odinger equation with $H_0$ Eq.(\ref{e:1}),
one gets the eigenenergies $E_{n,\ell}$ and the corresponding wave functions
$|n,\ell \rangle $ 
(to simplify the notations, we do not write explicitly the
dependence on magnetic quantum number $m$).
The 1S, 2S states correspond in our model to 
$J/\psi,~ \psi'$ respectively and 1P  to $\chi_c$ where
the different total spin state of $\chi_c $ are not split in energy.
(We consider that our $\chi_c$ is produced through a two-gluon
mechanism).

The time dependent
wave function for the $c\bar{c}$ in its rest frame is expanded on the basis
of eigenstates of $H_0$ Eq.(\ref{e:1}):
\be\label{e:5}
|c\bar{c},\ell \rangle (\tau)= \sum_{n=0}^{\infty}
c_{n,\ell}(\tau)e^{-{{i}}E_{n,\ell}{\tau }}|n,\ell \rangle .
\ee
($\hbar,c=1$). In practice, one truncates the sum to a number
 $\bar{n}$ of eigenstates, for a given $\ell$  
value. As we will discuss, we have chosen $\bar{n}$ large
enough so that the results we present are not sensitive to this truncation.

To model the interaction of the $c\bar{c}$ with the nuclear medium
 we add an imaginary part to Eq.(\ref{e:1}) : 
\be\label{e:2}
iW = i  {\gamma v \over 2} 
\sigma(\vec{r}_T,\sqrt{s_{\psi N}})
\rho(\vec{b},z)
\ee
where $v$ is the speed of the nucleus with respect to the
$c\bar{c}$ frame
and $\sigma$
 is the dipole cross section associated
to the interaction of the $c\bar{c}$ with a nucleon $N$,
$r_T$ is the transverse distance between $c$ and $\bar{c}$,
$\sqrt{s_{\psi N}}$ is the energy in the center of mass of the
$\psi N$ system and $\rho$ is the nuclear density evaluated at
the position $(\vec{b},z)$ of the $c\bar{c}$ center of mass, 
$z$ being the beam direction.
For the  dipole cross section 
we use the parametrization 
$\sigma(r_T,\sqrt{s})=\sigma_0(s)(1-e^{{-r^2_T}/{r_0^2(s)}})$,
determined by fitting deep inelastic scattering data and 
which well reproduces
the charmonium photoproduction data \cite{HK}. 
Concerning the nuclear density $\rho$, we present results obtained
with a Woods-Saxon profile,
with parameters 
chosen such as to well
reproduce the nuclear radii.

The time evolution of $c\bar{c}$ wavefunction (\ref{e:5}) 
in the nucleus is determined by
solving the time dependent Schr\"odinger equation for
$H=H_0-iW$ Eqs.(\ref{e:1}) and (\ref{e:2}).
This leads to a system of first-order coupled-channel 
differential equations for the amplitudes $c_{n,l}(t)$ :
\be\label{e:6}
\dot{c}_{n,\ell}(t)=-a \sum_{k=1}^{\bar{n}}c_{k,\ell}(t)
e^{{{i}}(E_{n,\ell}-E_{k,\ell}){t \over {\gamma}}}
\langle n,\ell |\sigma|k,\ell \rangle
\ee
with $a= v \rho(\vec{b},z)/2$. 
The time $t=\gamma \tau$ is now in the laboratory frame.
We have neglected in this first calculation 
the higher Fock states that should emerge from
the Lorentz boost \cite{HK}.
We see from (\ref{e:6}) that the imaginary potential (\ref{e:2})
introduces a transition amplitude among the charmonium eigenstates.

A difficult choice is that of the initial conditions $c_{n,\ell}(0)$
for (\ref{e:6}), related to the mechanism of hadroproduction of
charmonia, which is still badly known \cite{GH99,Vogt,Kramer,Peigne}.
We use:
\be\label{e:7}
\phi_{c\bar{c}, \ell} (0)= c_{\ell} f_{\ell}(r_T) 
e^{-{{1}\over{2}}\beta^2 (r_T^2+z^2)}
\ee
where $c_{\ell}$ is a normalization constant.
We describe the conversion of a gluon into a $c\bar{c}$ through
a gaussian multiplied by 
the function 
$f_{\ell}(r_T)=r_T^2$ (or $\vec{r}_T$) to account for 
the two (one) supplementary
gluons necessary to produce $J/\psi,\psi'$ ($\chi_c$). 
The initial wave function depends on one parameter only,
$\beta$. This parameter has been determined by fixing the ratio 
$|c_{2,0}/c_{1,0}|^2$ to the experimental ratio of $\psi'$
over $J/\psi$, produced in $pp$ collisions at 450~GeV, i.e. 
${B_{\psi' \rightarrow \mu \mu}\sigma^{pp \rightarrow \psi'}/ 
B_{J/\psi \rightarrow \mu \mu}\sigma^{pp \rightarrow J/\psi}}=(1.60\pm0.04)\%$  \cite{c2/c1}. 
where $B_{\psi \rightarrow \mu \mu}$ 
is the branching ratio to dimuon production and 
$\sigma^{pp \rightarrow \psi}$ the $pp$ reaction cross section.
This leads to the ratio $|c_{2,0}/c_{1,0}|^2(0)=0.22 \pm 0.03$
and $\beta=1.33 \pm 0.5$~GeV (the same $\beta$ has been used
for the $\chi_c$ states).

The number $\bar{n}$ of states in (\ref{e:5}),
to be included in the coupled channel equations (\ref{e:6}),
has to be large enough that the results 
do not depend on the truncation. 
We have 
checked how the results depend on the inclusion of extra states.
Up to 4 channels have been used for the S-states and up to 2 for
the P-states. The truncation always affects a little the last state
included; whereas the results for the lowest energy states are
practically unchanged. 

Let us now come to the results, obtained in infinite nuclear matter
first, i.e. by taking in (\ref{e:6}) a constant nuclear density $\rho=\rho_0$.
In fig.1 we show the time evolution of the probabilities associated to
$J/\psi$ and $\psi'$ production obtained by solving (\ref{e:6}) with
the initial conditions (\ref{e:7}), both neglecting and including the 
transition amplitudes 
between different eigenstates (non-diagonal terms). 
While in the former case the 
probabilities decrease exponentially, in the
latter  they present an oscillation. 
The interference pattern 
is more pronounced in the case of $\psi'$ whereas for $J/\psi$
the oscillations
stays very close to
the exponential. 
This effect directly influences the suppression
factors in $pA$ collisions as we will see.
The dominant oscillation frequency in Eq.(\ref{e:6})
is $\omega=(E_{2,\ell}-E_{1,\ell})/ \gamma$. 
We have also seen that 
the deviation from the uncoupled solution 
becomes stronger the 
higher $\gamma$ is.

Let us now consider a $c\bar{c}$ which is produced 
in a relativistic $pA$ collision
and evolves according to Eqs.(\ref{e:6}).
Integrating over all possible paths
the ratio 
$|{{c_{\psi}(\infty)}/{c_{\psi}(0)}}|^2$, which
gives the survival probability of the $\psi$ in the nucleus,
one gets the $pA \rightarrow
\psi$ reaction cross section : 

\be\label{e:9}
\sigma^{pA \rightarrow \psi}= \int d \vec{b} dz \rho(\vec{b},z)
\sigma^{pN \rightarrow \psi}
\Big| {{c_{\psi}(t(\vec{b},z))} \over
{c_{\psi}(0)}} \Big|^2
\ee
where $t(\vec{b},z)$ is the time necessary to a $c\bar{c}$
produced at a point $(\vec{b},z)$ to traverse the nucleus. 
If 
$\sigma^{pA \rightarrow \psi}=A\sigma^{pN \rightarrow \psi}$,
the suppression factor, defined as 
\be\label{e:0}
S^{\psi}_A \equiv {\sigma^{pA \rightarrow \psi} \over 
{A\sigma^{pN \rightarrow \psi} }}~,
\ee 
becomes  $S^{\psi}_A = 1$. 

Before making predictions, let us compare our calculations
to existing data. 
For $J/\psi$, we have to take into account
that the dimuons measured in the detector come from 
the decay of the directly produced $J/\psi$'s as well as those produced
by the decays of $\psi'$'s and $\chi_c$'s.
We will denote by $\mbox{``}J/\psi\mbox{''}$
the total number of produced $J/\psi$, i.e.
$S^{\mbox{``}J/\psi \mbox{''}}_{pA}=0.62S^{J/\psi}_{pA}+ 0.30S^{\chi_c}_{pA}+
0.08S^{\psi'}_{pA}$ \cite{pprecent} .

Let us first discuss charmonium production as a function of the
nuclear mass.
The $c\bar{c}$ pair travels along 
different lenghts, spending different time intervals
in the nuclei. Therefore, experiments with different $A$ explore 
the time dependence of the 
$c_{n,\ell}$ amplitudes (Fig.1).
We present results on $\mbox{``}J/\psi\mbox{''}$ and $\psi'$ (Fig.2)
in comparison with
the experimental values
by the NA50 Collaboration \cite{NA50},
obtained for $pA$ collisions with an impinging proton energy of $450~$GeV.
The data are given as branching ratio times cross section 
divided by the nuclear mass, which differs 
from the suppression by a constant. (For the comparison, we normalize
the data by this
constant determined from the 
average of the ratio experimental/calculated values.) 
We can see that our results are in good agreement with the experiment
both for $\mbox{``}J/\psi\mbox{''}$ and $\psi'$.
In Fig.3 we give predictions for the suppression of 
$\chi_c$ as a function of the nuclear mass, for different values of 
$\gamma$. The largest suppression is observed when $\gamma$
is the lowest, corresponding to a longer time spent 
in the nucleus.

Let us now look at the $x_F$ dependence. 
The suppression factor is often parametrized 
as  $S^{\psi}_A(x_F,\sqrt{s_{pp}}) = A^{\alpha_{\psi}(x_F,\sqrt{s_{pp}},A)-1}$
\cite{GH99}. 
In fig.4  
the $\alpha$  values
for $\mbox{``}J/\psi\mbox{''}$ and $\psi'$ are
compared
to experimental data on $pA$ collisions, 
obtained by the E866/NuSea Collaboration \cite{E866}.
The impinging proton energy is 800~GeV and the label A is the ratio
of the atomic numbers of tungsten over beryllium.
In this region of small positive and negative $x_F$ values,
our calculations 
are in good agreement with the
experimental values for $\mbox{``}J/\psi\mbox{''}$, 
whereas for $\psi'$ they overpredict the
data. In fact, 
in the region of small positive $x_F$, 
the $c\bar{c}$  can traverse the nucleus before developing
into a $\psi$ meson. In this kinematical regime,
other effects missing in our model, like for example 
gluon shadowing \cite{Bor},
can play a role.

Finally, we present  predictions for the $\alpha$ values (Fig.5)
and for the suppression factors (Fig.6) for
$\mbox{``}J/\psi\mbox{''},\psi'$ and $\chi_c$
in the negative $x_F$ region, 
where
experimental data from the
HERA-B Collaboration \cite{HERAB}
at DESY and the PHENIX Collaboration at RHIC \cite{PHENIX}
will soon be available.
As we can see, both $\mbox{``}J/\psi\mbox{''}$ 
and $\chi_c$ present
a rather flat behavior,
whereas  $\psi'$ is very sensitive to the $x_F$ values.
Besides, the results on $\psi'$  also show
a stronger variation with the mass of the nucleus.
This can be understood from the fact that
the effect of the coupling between the charmonium eigenstates is much 
stronger for $\psi'$ than for $J/\psi$ 
(and $\chi_c$), as one can see from Fig.1.

Fig.5 also shows that 
the $\alpha$ parametrization 
of the suppression (often
used with $\alpha$ independent of $A$) is especially not good
in the negative $x_F$ region, since there is
a strong dependence of the $\alpha$ values 
on the nuclei. 

In summary, we have studied the production of charmonium
($J/\psi,~\psi',~\chi_c$) in relativistic proton-nucleus collisions.
We focussed on the nuclear effects on the $c\bar{c}$ 
time evolution and on the charmonium production when 
the time for a 
$c\bar{c}$ to develop to a $\psi$ meson 
is short compared to the time necessary to traverse the nucleus. 
This kinematical region may be explored experimentally 
by looking at slow enough
$c\bar{c}$ and therefore charmonium produced in the negative $x_F$
region.
The quantum-mechanical model used here includes
a realistic potential binding the $c\bar{c}$ pair and an imaginary
potential, describing 
the collision of the $c\bar{c}$ with the nucleons and
depending on the dipole charmonium-nucleon cross section.
The imaginary potential introduces transition amplitudes among
charmonium eigenstates.
We have shown that this produces an interference pattern 
on the charmonium survival probability, particularly important for 
$\psi'$, and affects the charmonium suppression factors. 
The results are compared to experimental data from the NA50
and E866/NuSea Collaborations.
We have made predictions on the suppression factors
of $J/\psi$, $\psi'$ and
$\chi_c$,  for different nuclei 
in the fragmentation region of the nucleus, where
experiments will be performed soon 
by the
HERA-B Collaboration 
at DESY and the PHENIX Collaboration at RHIC. We expect that 
the exploration of the negative $x_F$ region will
bring new information on the charmonium-nucleon cross
section and on the initial conditions which are not well
known, especially for $\chi_c$.
A better understanding of the mechanisms for charmonium production 
in proton-nucleus collisions in this yet unexplored region
will help us in the interpretation
of the measurements on relativistic nucleus-nucleus collisions.

\vspace{0.3cm}
\noindent

We are very grateful to J\"org H\"ufner
for having interested us in this problem. 
We thank him for 
his useful suggestions and comments all along the
realization of this work, and for his very careful reading of the
manuscript. 
We thank Boris  Kopeliovich and Stephane Peign\'e
for very useful discussions and comments, as well as Yuri Yvanov for
his kind help.  
One of us (C.V.) acknowledge all the people at the
Institut f\"ur Theoretische Physik  
of the University of Heidelberg, for the very warm hospitality during her stay.

\begin{figure}
\begin{center}
\includegraphics[angle=0.,scale=0.35]{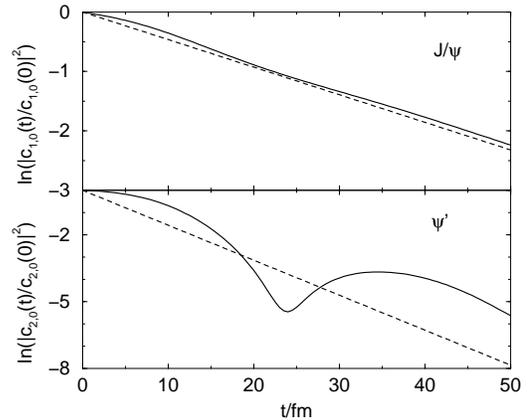}
\end{center}
\protect\caption{Probability of surviving in the nuclear medium 
of a $J/\psi$ (upper)
and a $\psi'$ (lower) as a function of time (in the laboratory
frame), normalized to the initial probability. The transition
amplitudes among
different charmonium states, due to the medium, 
produces an oscillatory behaviour (solid
line) whereas the solution is simply an exponential (dashed line)
when they are absent.}
\end{figure}

\begin{figure}
\begin{center}
\includegraphics[angle=0.,scale=0.35]{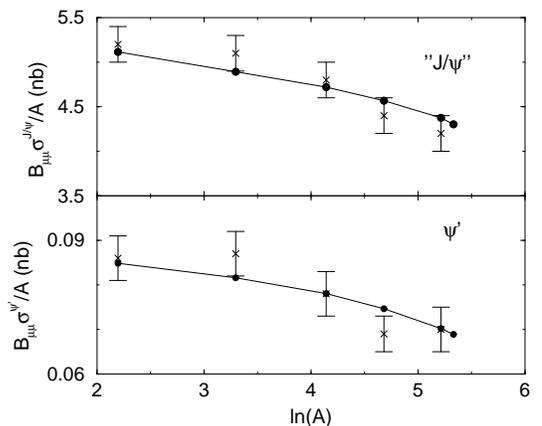}
\end{center}
\protect\caption{Results for the $\psi$ production cross section
times the branching ratio to dimuon production, $B_{\mu \mu}\sigma^{\psi}$,
for the
  $\mbox{``}J/\psi\mbox{''}$ (top) and $\psi'$ (bottom)
produced in $pA$ collisions with $E_p=450~$GeV
 as a function of the nuclear mass. 
The experimental data obtained by the NA50 Collaboration [7]
are shown for comparison.
(The curves are only given to guide the eye).} 
\end{figure}

\begin{figure}
\begin{center}
\includegraphics[angle=0.,scale=0.4]{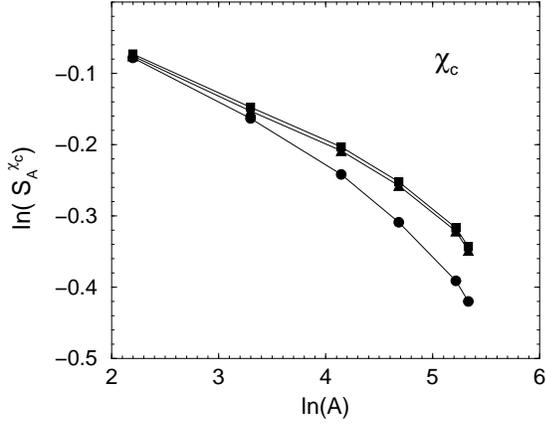}
\end{center}
\protect\caption{Predictions on the suppression factor
of the $\chi_c$ meson, produced in relativistic
$pA$ collisions, as a function of
  the mass of the nucleus $A$. We present results for 
$\gamma=5$~(circles), $16$~(squares), $21$~(triangles).}
\end{figure}

\begin{figure}
\begin{center}
\includegraphics[angle=0.,scale=0.35]{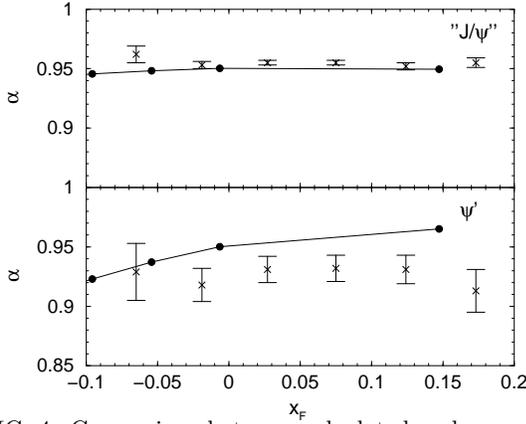}
\end{center}
\protect\caption{
Comparison between calculated and experimental $\alpha$ values for
$\mbox{``}J/\psi\mbox{''}$ (top) and $\psi'$ (bottom)
as a function of Feynman $x_F$ obtained for $pA$ collisions with 
$E_p=800$~GeV. Here the $\alpha$ values are relative to the
ratio $W/Be$. The data are from
the E866/NuSea Collaboration [8].}
\end{figure}

\begin{figure}
\begin{center}
\includegraphics[angle=0.,scale=0.45]{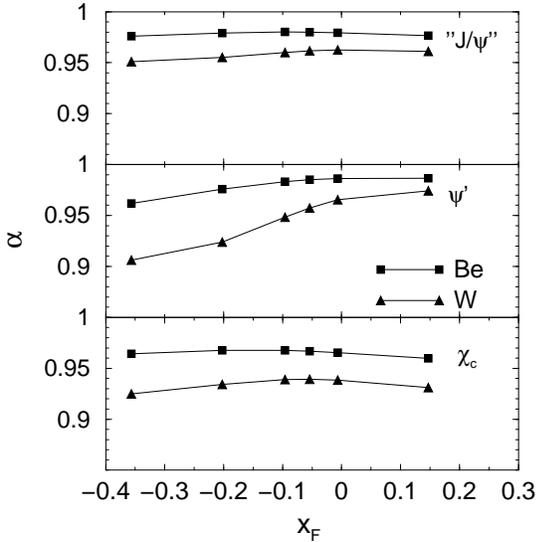}
\end{center}
\protect\caption{Predictions for $\mbox{``}J/\psi\mbox{''}$ 
(top), $\psi'$ (middle)
and $\chi_c$ (bottom),
on the dependence of the
$\alpha$ values as a function of $x_F$,
obtained with $E_p=800$~GeV.
We present results obtained for tungsten (triangles) and beryllium (squares). }
\end{figure}

\begin{figure}
\begin{center}
\includegraphics[angle=0.,scale=0.4]{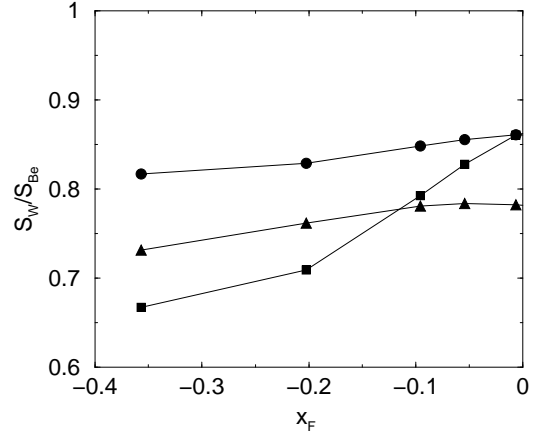}
\end{center}
\protect\caption{Predictions for 
the suppression factors for $\mbox{``}J/\psi\mbox{''}$ (circles), $\psi'$ (squares)
and $\chi_c$ (triangles)
as a function of the negative $x_F$ region,
obtained in a $pA$ collision with $E_p=800$~GeV and $A$ being the ratio
of tungsten over beryllium.}
\end{figure}

\end{document}